# The solar towers of Chankillo

**Amelia Carolina Sparavigna**
Department of Applied Science and Technology
Politecnico di Torino, C.so Duca degli Abruzzi 24, Torino, Italy

*An ancient solar observatory is composed by thirteen towers lined on a hill of a coastal desert of Peru. This is the Chankillo observatory. Here we discuss it, showing some simulations of the local sun direction. An analysis of the behaviour of shadows is also proposed.*

One of the oldest solar observatories in America is probably that in the archaeological heritage area around Chankillo, in the Casma-Sechin oasis of the Peruvian coastal desert. It is a more than 2,300 years old site, that unfortunately, as the press is reporting, is threatened because of some works planned by agricultural companies [1]. The fact that a more or less straight line of towers built on a hill could be a solar observatory came from the researches lead by I. Ghezzi, a member of the Institute of Archaeological Research (IDARQ), and C. Ruggles of the University of Leicester [2-4]. In April 2008, Chankillo was declared Cultural Patrimony of the Nation by the National Institute of Culture (INC) of Peru.

The monumental complex of Chankillo is including the hilltop fort, the Thirteen Towers solar observatory, and a large residential and gathering area. Archaeologists have nicknamed the hilltop fort the "Norelco ruin" (see Fig.1), because of its resemblance to a modern electric shaver [5]. This structure has thick walls and the location on the top of a hill suggested it was a fort, but some facts puzzled the researchers because this fort had many gates and no water sources inside. May be it was a worship place.

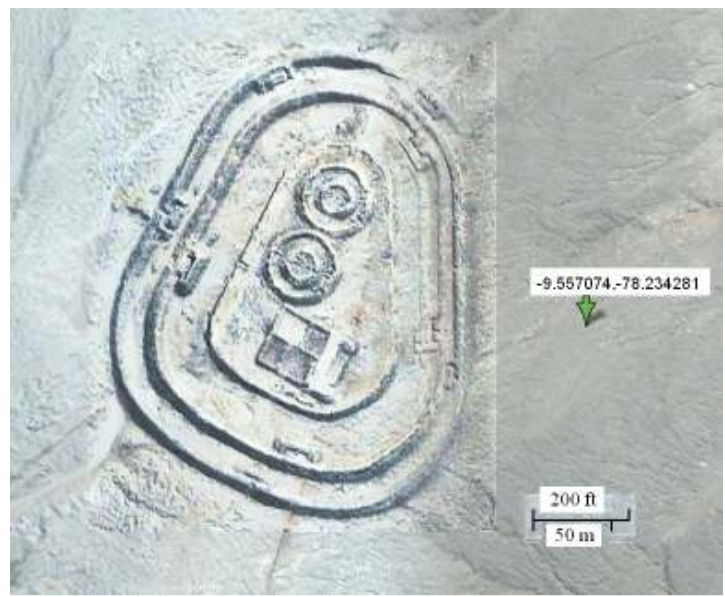

Fig.1 The Chankillo Fort as obtained after a processing of an image of Google Maps.

Southeast of the central complex we find the Thirteen Towers, which resemble the vertebrae of a backbone. These towers run on a hill separated by about 5 meters. As Ref.5 is showing, on either side of this line of towers are guessed the presence of observing points, but little is left of the eastern observation structure. South of the eastern observing point there is another building complex, probably used for food storage.

The Thirteen Towers and the two observing points can be used as a solar observatory.

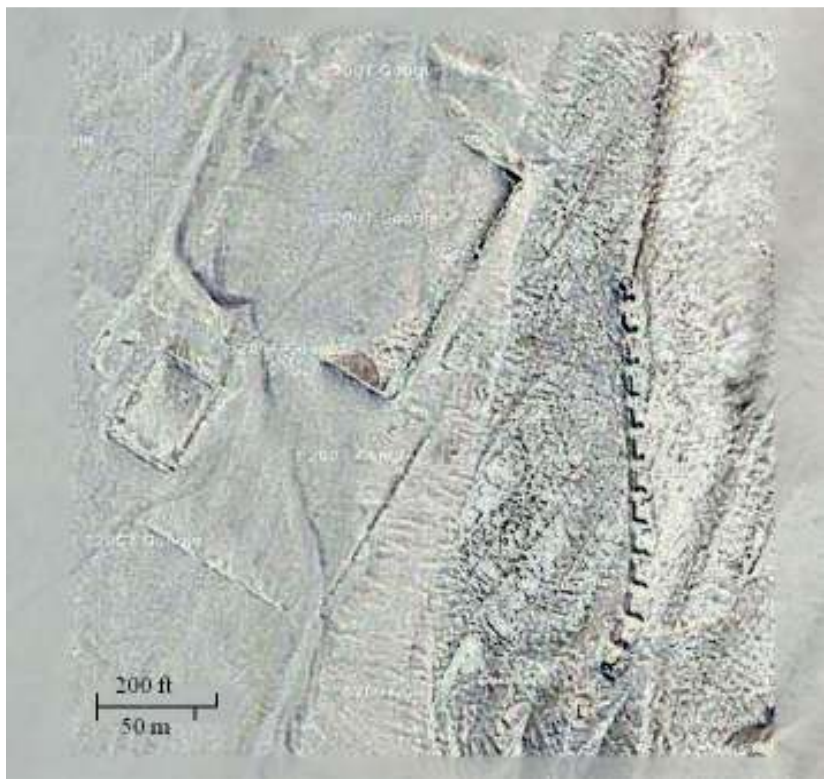

Fig.2 The towers of Chankillo look like the vertebrae of a backbone.

As it is told in Ref.5, the observing points are situated so that, on the solstices, the sunrises and sunsets line up with the towers at either end of the line, suggesting that this ancient civilization had a solar calendar. Wikipedia [6] remarks that, in addition to potential ceremonial purposes, "the observatory may have had practical uses as well. In Peru's dry coastal reason, precipitation is seasonal, so a reliable solar calendar would help determine the optimal time to plant crops."
For European people, the prototypes of ancient solar observatories are the sites of Stonehenge and Goseck [7]. The European sites have generally a circular structure, but, as we have seen Chankillo is composed of a more or less straight line of towers. This structure is more suitable for the observation of sunrise, sunset and zenith position of the sun. Let us remember that the location is in the tropical zone and therefore the sun reaches the zenith two times during the year.
Quite recently [8], I have discussed the orientation of the "wheels" of the Syrian desert, that is of some Neolithic stone circles with radial lines that can be found in this desert, comparing them with the directions of sunset, noon, and sunrise on solstices. To this purpose I used a software developed for solar energy applications, available at the site http://www.sollumis.com/. This web site is providing a model of sunlight direction into a Google map on any day of the year. In the Appendix, the Figure 1 of Ref.8 is proposed to show one of the stone circles of the Syrian desert analysed with the mentioned software by sollumis.com.
For the sun towers of Chankillo, the simulation is given in the Fig.3: we see that on June 21, the sun rises northern the first tower and that on December 21, the sun is rising southern the last tower.
Besides the sunrise and sunset, we could ask ourselves what are the other phenomena that the ancient people of Chankillo could have observed. One is the zenith position of the sun, because as previously told, Chankillo is in the tropical zone, and, as remarked in [9], it was an event quite important for the ancient Central and South American cultures. Moreover there is the behaviour of the shadows. This behaviour was well known to ancient people, as Ptolemy tells in his Almagest (Appendix B discusses the behaviour of the shadows cast by gnomons at Napata, Sudan).

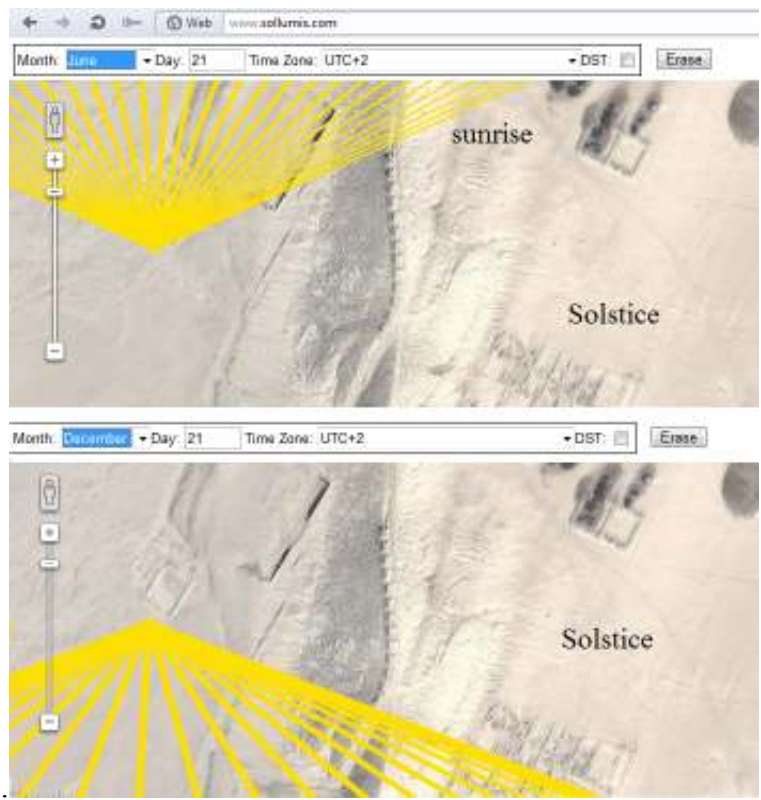

Fig.3 The direction of the sun as observed from the western observation point, simulated using sollumis.com software. In the upper image we see that on the June solstice, the sun rises northern the first towers. In the lower panel, the sun on the December solstice is rising southern the last tower. The lines show the direction and height (altitude) of the sun. Thicker and shorter lines mean the sun is higher in the sky. Longer and thinner lines mean the sun is closer to the horizon.

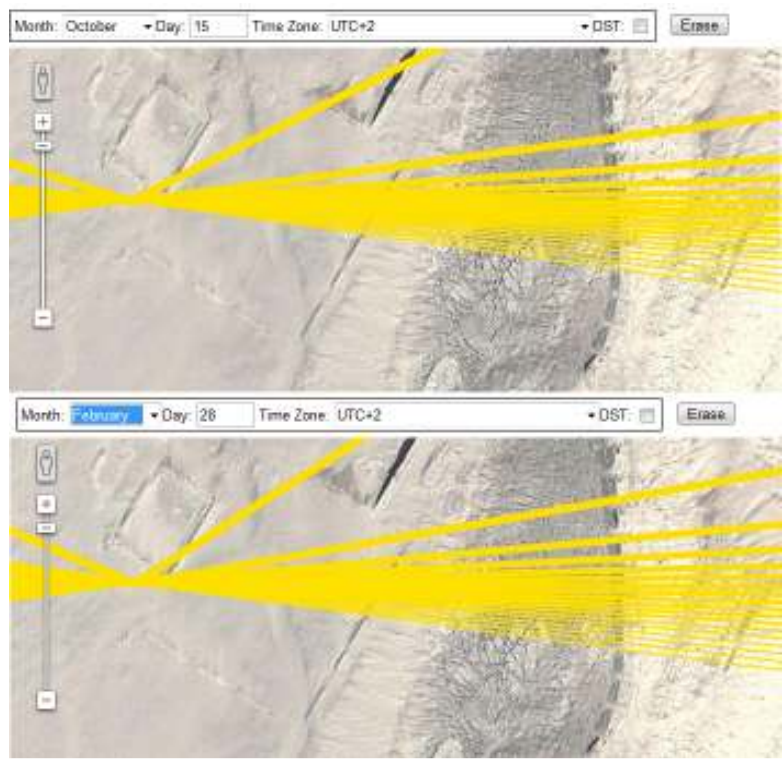

Fig.4 The shadows change direction. In the morning and evening the shadow has a component towards North, whereas during the day, this component is opposite, that is towards South. The images correspond to October 15 and February 28. October 18 and February 25 have the noon altitude at 90°.

As we can see. in Chankillo, the shadows are pointing towards South from the equinox of March 21 to the equinox of September 21. From October 18 till February 24, the shadows are pointing towards North. In the other days, the shadows change direction during the day That is, shadows have a component pointing towards South about noon, whereas they are pointing towards North in the morning and in the evening. Two examples are shown in Fig.4, with simulations. From sollumis.com, we find that on October 18 and February 25 we have the noon altitude at 90°. It seems that there is some difference with data of Ref.3.

Besides the investigation of some correspondence of the sunrise and sunset between towers and the modern calendar, as proposed in Ref.3, it could be interesting a detailed study of the behaviour of shadows of the towers Since the observation of shadows is quite old, much older than the Ptolemy's words, we can argue that the people of Chankillo knew very well this behaviour too. May be, the inhabitants of Chankillo observed carefully the shadows of the towers, and combining their behaviour with the direction of sunrise and sunset, they could forecast, as Wikipedia remarks, the seasonal precipitations and determine the optimal period for planting crops.

**Appendix A**

For the reader convenience, an image from Ref.8 is proposed. It is showing one of the stone structure of Syrian Desert. The image shows the directions of sun during the day. "The lines on the drawing show the direction and height (altitude) of the sun throughout the day. Thicker and shorter lines mean the sun is higher in the sky. Longer and thinner lines mean the sun is closer to the horizon", according to Sollumis.com http://www.sollumis.com/. On the left, the site as it appears in the Google Maps. In the middle, the direction of the sun on the summer solstice, choosing the center of the circle for observation. We see that, at sunrise, the sun is passing near the dot. At the sunset the direction is that of a line. In the image on the right, we see the direction of the sun on winter solstice. At sunrise, the lines is passing between dots. The sunset has the direction of a radius of the stone structure.

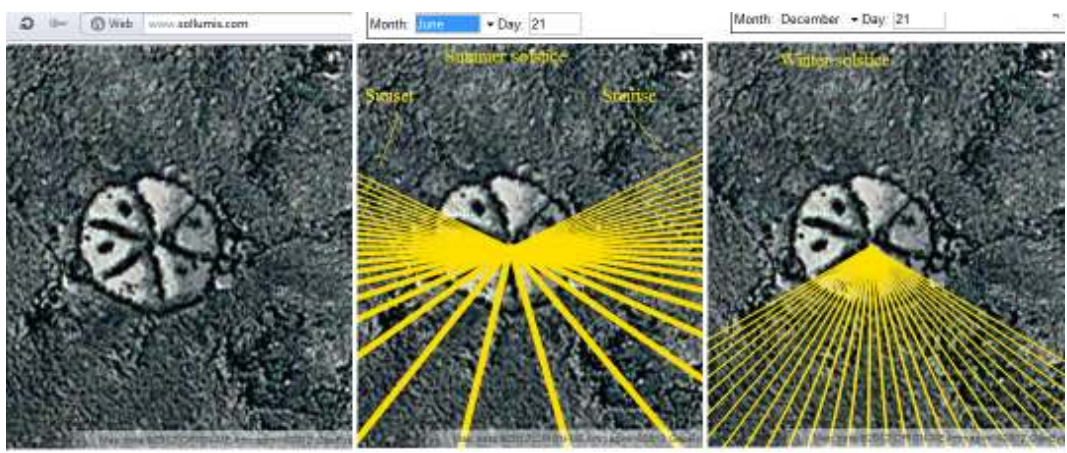

Fig.1 of the paper entitled

**Appendix B**

In the Almagest written by Ptolemy, it is discussed the shadow of a gnomon at the latitude of the ancient Napata, in Sudan, in the second book, chapter 6. He is telling that there is passing the sixth parallel, with a longest day of 13+1/4 equinoctial hours [10]. In Ref.10 it is told that this town was 20°14' from the equator (in fact, Napata was close the modern Karima, 18° 33′ 0″ N, 31° 51′ 0″ E). This is one of the parallels with the shadow of the gnomons going both way and where the sun comes into the zenith twice a year, and makes the gnomons shadowless at noon, when it is 31° from the summer solstice. Then, when the sun is at the zenith of a site having a latitude of 20°14',

Ptolemy from his tables was able to tell the sun being about 31° from Solstice (31 days before the solstice, that is 20 of May). For 62 days about the summer solstice, the gnomon casts a shadow pointing towards South (South-East, South-West) for all the day. Then, during a certain period of the year, the shadow is pointing towards North about noon, whereas it is pointing towards South in the morning and in the evening. From the autumn equinox to the spring equinox, the shadows points towards North. We can easily check this behaviour using sollumis.com.

This fact was well-known by people that had to travel in tropical regions. An English mathematician/astronomer Thomas Keith (1759-1824) writes [11], not so differently from Ptolemy, that "If a horizontal dial, which shows the hour by the top of the perpendicular gnomon, be made for a place in the torrid zone, whenever the sun's declination exceeds the latitude of the place, the shadow of the gnomon will go back twice in the day, once in the forenoon and once in the afternoon, and the greater the difference between the latitude and the sun's declination is, the farther the shadow will go back."